\documentclass[aps,prl,amsmath,amssymb,twocolumn,notitlepage]{revtex4-1}

\usepackage{amsmath}
\usepackage{amssymb}
\usepackage{bm}
\usepackage{hyperref}

\usepackage{xcolor}
\usepackage{braket}
\usepackage{graphicx}

\newcommand{\ra}{\rangle}
\newcommand{\la}{\langle}

\newcommand{\dm}{\mathrm{d}}
\newcommand{\ov}{\overline}

\newcommand{\tr}{{\rm Tr}}

\def\bk{{\bm{k}}}
\def\BZ{{\rm{BZ}}}
\def\sgn{{\mathrm{sgn}}}
\def\tbk{{\bm{\tilde k}}}
\def\bR{{\bm{R}}}

\def\bA{{\bm{A}}}

\def\bnabla{{\bm{\nabla}}}

\def\bq{{\bm{q}}}

\def\br{{\bm{r}}}
\def\Tr{{\rm{Tr}}}
\def\bb{{\bm{b}}}
\def\tBe{{\tilde B^{(\epsilon)}_i}}
\def\mA{{\mathcal A}}
\def\bmA{{\bm{\mA}}}
\DeclareMathOperator\supp{supp}

\begin{document}

\title{Constraints on real space representations of Chern bands}
\author{Qingchen Li}
\affiliation{Department of Physics, Harvard University, Cambridge, MA 02138, USA}
\author{Junkai Dong}
\affiliation{Department of Physics, Harvard University, Cambridge, MA 02138, USA}
\author{Patrick J. Ledwith}
\affiliation{Department of Physics, Harvard University, Cambridge, MA 02138, USA}
\author{Eslam Khalaf}
\affiliation{Department of Physics, Harvard University, Cambridge, MA 02138, USA}
\date{\today}
\begin{abstract}
A Chern band is characterized by a Wannier obstruction indicating the absence of a basis of complete, orthogonal, and exponentially-localized states. 
Here, we study the properties of real space bases of a Chern band obtained by relaxing either exponential localization or orthogonality and completeness. This yields two distinct real space representations of a band with Chern number $C$: (i) a basis of complete orthogonal Wannier states which decay as power-laws and (ii) a basis of exponentially-localized overcomplete non-orthogonal coherent states. For (i), we show that the power-law tail only depends on the Chern number and provide an explicit gauge choice leading to the  universal asymptotic $w(\br) \approx \frac{C e^{-i C \varphi_\br}}{2\pi |\br|^2}$ up to a normalized Bloch-periodic spinor. 
For (ii), we prove a rigorous lower bound on the spatial spread that can always be saturated for ideal bands. We provide an explicit construction of the maximally localized coherent state by mapping the problem to a dual Landau level problem where the Berry curvature and trace of quantum metric take the roles of an effective magnetic field and scalar potential, respectively. Our coherent state result rigorously bounds the spatial spread of {\it any} localized state constructed as a linear superposition of wavefunctions within the Chern band. Remarkably, we find that such bound does not generically scale with the Chern number and provide an explicit example of an exponentially localized state in a Chern $C$ band whose size does not increase with $|C|$. Our results show that band topology can be encoded in a real space description and set the stage for a systematic study of interaction effects in topological bands in real space.
\end{abstract}
\maketitle

\emph{Introduction}--- The discovery of band topology had a transformative impact on the study of quantum materials~\cite{HasanKane, MooreTI, QiZhang}. While early work on topological bands focused on their quantized response and
surface states, more recent work have centered the notion of Wannier obstructions, the absence of an orthonormal basis of exponentially localized orbitals describing the band ~\cite{ThonhauserVanderbilt, SoluyanovVanderbilt, Wannierizability, po2017symmetry, bradlyn2017topological, FragileTopology, ShiftInsulator}. 
The discovery of topological flat bands in moir\'e materials have further emphasized the 
importance of Wannier obstructions to understand strong correlations 
in these systems \cite{ahn2019failure, Po2019faithful, song2021twisted}, distinguishing platforms where a Hubbard description is possible from those where more complicated descriptions in terms of flat Chern bands~\cite{Tarnopolsky2019origin, BultinckHidden, ledwith2021strong} or topological heavy fermions~\cite{song2022TBGTHF,calugaru2023TBGTHF2,yu2023TTGTHF,herzogarbeitman2024topologicalheavyfermionprinciple} are more appropriate.

Despite its central importance, the concept of Wannier obstruction has so far been studied as a binary distinction; two bands with non-zero but different Chern numbers differ in several fundamental ways but the notion of Wannier obstructions places them in the same category. 
Moreover, despite the absence of an exponentially localized Wannier basis, several physical questions related to interaction physics~\cite{zang2022real, huang2024quantum} and bound states 
~\cite{queiroz2024ring, Lawflat, Lawanomalous} in a topological band are naturally phrased in terms of a real space basis.

\begin{figure*}
    \centering
    \includegraphics[width = \textwidth]{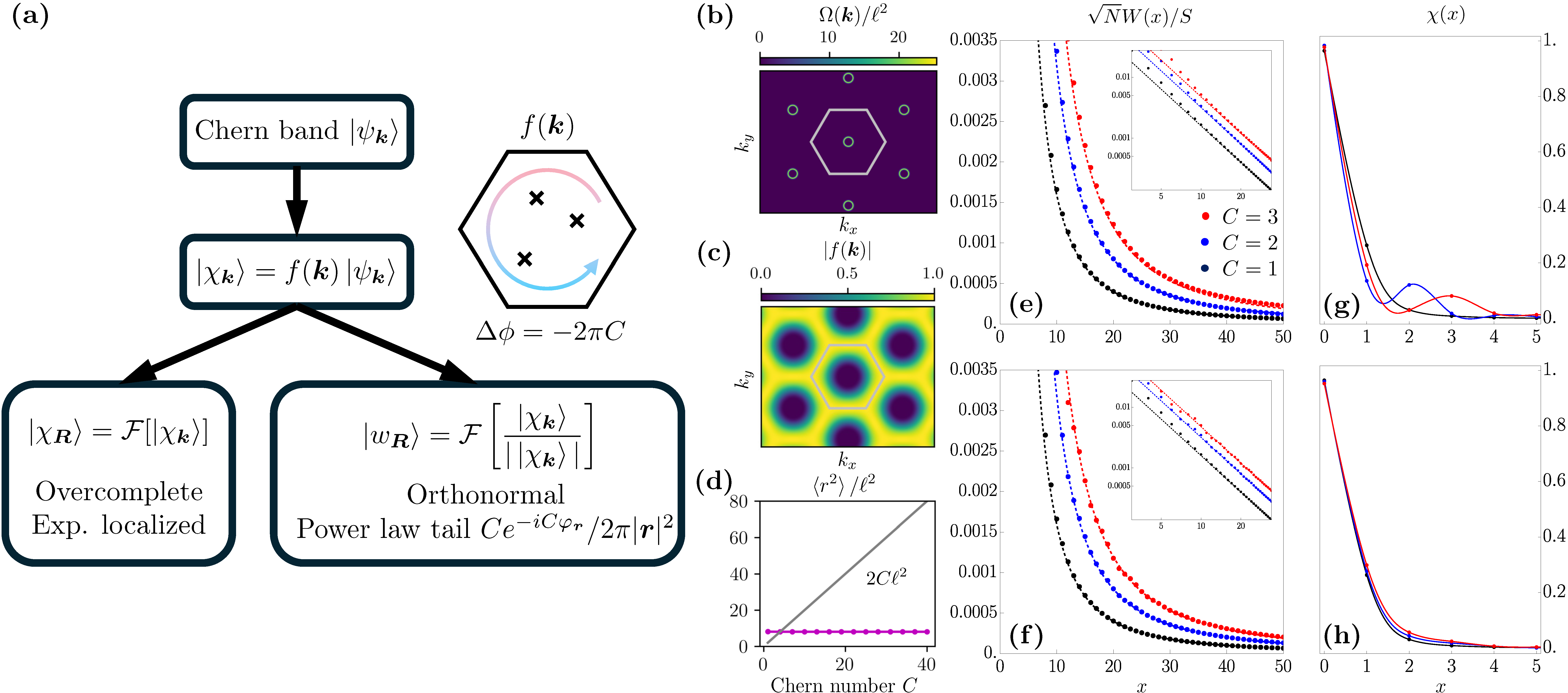}
    \caption{(a) Real space bases in a Chern band: exponentially localized but overcomplete coherent states or power-law orthonormal Wannier states. The Chern number is related to the phase winding of the smooth function $f(\bk)$. 
    (b) Concentrated Berry curvature distribution for the model (\ref{higherCmodel}) with $s=0.2$ and $C=8$. (c-d) Coherent states construction for the model (\ref{higherCmodel}): by choosing $f(\bk)$ that avoids the concentrated Berry curvature, we construct localized states whose spatial spread $\braket{r^2}$ does not grow with the Chern number $C$. 
    Wannier states and coherent states for the Hamiltonians (\ref{HD1}) (e,g) and (\ref{HD2}) (f,h). The dashed lines in panels (e) and (f) indicate the Wannier asymptotics (Eq.~\ref{WA1}). For the latter, since we are studying the asymptotics along the $x$ direction and the zeros lie along the $x = 0$ line, we have $\phi(x) = \frac{1}{|\sqrt{C}|} \sum_{i=1}^{|C|} 1 = |\sqrt{C}|$.}
    \label{fig:Fig1}
\end{figure*}

Motivated by these consideration, in this letter, we prove several rigorous constraints on any real space basis describing a Chern band focusing on two classes of bases obtained by relaxing either the orthonormality or the exponential localization of Wannier functions. The former leads to an exponentially localized but overcomplete coherent state basis whereas the latter leads to a power-law localized orthonormal Wannier basis.  
Our work establishes a novel way to characterize topology in terms of real space bases and provides rigorous limitations on such bases with important implications to models of interacting topological bands. Furthermore, our construction of coherent states imposes constraints on {\it any} localized state formed in a topological band. Our approach, summarized in Fig.~\ref{fig:Fig1}(a), constructs an overlap function $f(\bk)$ whose zeroes are related on one hand to the Chern number and on the other hand to the asymptotics and spatial spread of Wannier and coherent state bases, respectively. 

\emph{Wannier functions and the projection method}--- We start by reviewing the formalism of Wannier functions (WFs)~\cite{marzari2012maximally}. For simplicity, we will focus on the single band case 
and discuss multiband generalization later. A WF $|W_\bR \rangle$ labelled by a lattice site $\bR$ is constructed via the Fourier transform of the Bloch wavefunction $|\psi_\bk \rangle$; $|W_\bR \rangle = \frac{1}{\sqrt{N}} \sum_{\bk \in \BZ} e^{-i \bk \cdot \bR} |\psi_\bk \rangle$, where $N$ is the number of lattice sites. WFs form a complete orthonormal basis for the band, but are not unique due to the gauge ambiguity of Bloch functions, 
$|\psi_\bk \rangle \mapsto e^{i \theta(\bk)} |\psi_\bk \rangle$. In the real space representation $|\br \rangle$~\cite{footnote3} 
, the WFs $W_\bR(\br) := \langle \br|W_\bR \rangle$ satisfy $W_\bR(\br) = W(\br - \bR)$, i.e. the WF labelled by $\bR$ is centered at the position $\br = \bR$. In a gauge where $|\psi_\bk \rangle$ is a smooth and periodic function of $\bk$ under reciprocal lattice translations, the corresponding WFs $W(\br)$ are exponentially localized in $\br$. A band is called Wannierizable if there exists such exponentially localized set of WFs. Without extra symmetry restrictions, 
a 2D band is Wannierizable iff its Chern number is 0~\cite{Wannierizability}. Our goal now is to study the properties of real space bases in Chern bands, obtained by relaxing one of the two main properties of WFs, orthogonality or exponential localization.

Our approach generalizes the construction of real space bases in the Landau levels~\cite{Rashba, BargmannCoherentState,PerelomovCompleteness} to Chern bands by identifying how the standard approach to construct WFs, called the projection method~\cite{WannierRMP}, fails. In the projection method, we start from a trial set of orbitals $|\tau_\bR \rangle$, whose Fourier transform is $|\tau_\bk \rangle := \frac{1}{\sqrt{N}} \sum_{\bR} e^{i \bk \cdot \bR} |\tau_\bR \rangle$. We then project $|\tau_\bk \rangle$ on our band via $|\chi_\bk \rangle = P_\bk |\tau_\bk \rangle$, where $P_\bk := |\psi_\bk \rangle \langle \psi_\bk|$ is the band projector at momentum $\bk$. Crucially, $P_\bk$ is smooth and periodic in $\bk$ for any isolated band (even for non-zero Chern number). Similarly, $\ket{\tau_\bk}$ is smooth and periodic since $|\tau_\bR \rangle$ is a local basis. As a result, $|\chi_\bk \rangle$ is smooth and periodic, but not necessarily normalized. Its norm is given by $\langle \chi_\bk|\chi_\bk \rangle = |f(\bk)|^2$ where $f(\bk) := \langle \psi_\bk|\tau_\bk \rangle$. If $f(\bk)$ is non-zero everywhere, we can define a smooth, periodic, and normalized basis $|\tilde \psi_\bk \rangle := \frac{|\chi_\bk \rangle}{|f(\bk)|}$ whose Fourier transform yields orthonormal exponentially localized WFs. The phase factor $e^{i \theta(\bk)} = \frac{f(\bk)}{|f(\bk)|}$ is a gauge transformation which makes $|\psi_\bk \rangle$ smooth and periodic since $|\tilde \psi_\bk \rangle = \frac{P_\bk |\tau_\bk \rangle}{|f(\bk)|} = \frac{f(\bk)}{|f(\bk)|} |\psi_\bk \rangle$. Thus, a band is Wannierizable iff we can find a basis $|\tau_\bk \rangle$ such that $f(\bk) = \langle \psi_\bk|\tau_\bk \rangle$ is non-zero for all $\bk$~\cite{footnote1}

In a Chern band where $f(\bk)$ has zeroes, we can define two types of real space bases. Taking the Fourier transform of the smooth, periodic but unnormalized state $|\chi_\bk \rangle = f(\bk) |\psi_\bk \rangle$ yields an exponentially localized but overcomplete basis $|\chi_\bR \rangle$, which we will call coherent states, that cannot be made orthonormal in the thermodynamic limit. In contrast, taking the Fourier transform of $|\tilde \psi_\bk \rangle = \frac{|\chi_\bk \rangle}{|f(\bk)|}$ yields an orthonormal Wannier basis that decays as a power-law due to the zeros of $f(\bk)$. Our construction proceeds by first establishing a rigorous connection between the zeros of $f(\bk)$ and the Chern number, then showing that these zeros control the power-law tails of the Wannier function and the properties of the overcomplete coherent states.

\emph{Chern number from wavefunction singularities}--- To relate the Chern number to the zeros of $f(\bk)$, consider a smooth non-periodic gauge for $|\psi_\bk \rangle$. An alternative proof using a periodic singular gauge is provided in the Supplemental Material (SM)~\cite{SM}. 
For a Chern band, a smooth Bloch wavefunction cannot be periodic in $\bk$ and instead satisfies $|\psi_{\bk + \bb_l} \rangle = e^{i \theta_l(\bk)} |\psi_{\bk} \rangle$, $l = 1,2$, where $\bb_1$ and $\bb_2$ are reciprocal lattice vectors~\cite{footnote2}. 
To obtain a consistent phase for $|\psi_{\bk + \bb_1 + \bb_2} \rangle$, the combination $\theta_2(\bk + \bb_1) + \theta_1(\bk) - \theta_1(\bk + \bb_2) - \theta_2(\bk)$ has to be an integer multiple of $2\pi$, which we will show to be equal to the Chern number. First note that the Berry connection $\bA(\bk) := i\langle u_\bk|{\boldsymbol \nabla}_\bk|u_\bk \rangle$ satisfies $\bA(\bk + \bb_l) = \bA(\bk) - {\boldsymbol \nabla}_\bk \theta_l(\bk)$. Furthermore, $|\tau_\bk \rangle$ is smooth and periodic which means that $f(\bk + \bb_l) = e^{-i \theta_l(\bk)} f(\bk)$. This can be solved for $\theta_l(\bk)$ yielding $\theta_l(\bk) = i [\ln f(\bk + \bb_l) - \ln f(\bk)]$ which implies $\bA(\bk + \bb_l) - \bA(\bk) = -i{\boldsymbol \nabla}_\bk [\ln f(\bk + \bb_l) - \ln f(\bk)]$. The Chern number is the integral of the Berry connection around the BZ. The contribution to this integral from two opposite sides of the BZ, let's say those parallel to $\bb_1$, can be written as $\sim \int_0^1 dk_1 \bb_1 \cdot [\bA(k_1 \bb_1 + \bb_2) - \bA(k_1 \bb_1)] = -i\int_0^1 dk_1 \bb_1 \cdot {\boldsymbol \nabla}_\bk [\ln f(k_1 \bb_1 + \bb_2) - \ln f(k_1 \bb_1)]$, leading to 
\begin{equation}
    C = -\frac{1}{2\pi i}\oint d\bk \cdot {\boldsymbol \nabla}_\bk \ln f(\bk)
    \label{Chernfk}
\end{equation}
Hence, the Chern number is given by the phase winding of $f(\bk)$. Since $f$ is smooth, its phase winding is given by the phase winding around its zeros. For the $i$-th zero of order $m_i$ at $\bk = \bk_i$, we can write $f(\bk_i + \bq) \approx \prod_{l=1}^{m_i} (\alpha^i_l q + \beta^i_l \bar q)$ where $q = q_x + i q_y$ and $\alpha^i_l$ and $\beta^i_l$ are complex constants with $|\alpha^i_l| \neq |\beta^i_l|$~\cite{SM}. The winding of the $i$-th zero, computed from Eq.~\ref{Chernfk}, is $\delta_i = \sum_{l=1}^{m_i} \sgn(|\alpha^i_l| - |\beta^i_l|)$. Thus, the Chern number is given by the total winding around zeros, $C = -\sum_i \delta_i$.

By choosing different trial states, we get different functions $f(\bk)$ with different zeros and different behavior around those zeroes but the total winding of zeros remains the same. In SM~\cite{SM}, we prove that it is always possible to choose the trial states $|\tau_\bk \rangle$ to satisfy the following: (i) redundant zeros with no winding, e.g. $f(\bk_0 + \bq) \sim q \bar q$, can always be removed. This means that an $m$-th order zero will have winding $\pm m$ (ii) $f(\bk)$ behaves as $\propto (q_x + i \sgn(\delta_i) q_y)^{|\delta_i|}$ around the $i$-th zero, and (iii) zeros can be moved anywhere in the BZ, two zeroes of winding $\delta_i$ and $\delta_j$ can be combined in a single zero of winding $\delta_i + \delta_j$ and vice versa. 
We will find it useful to consider two special gauge choices:  (i) single-zero (SZ) gauge where $f(\bk)$ has a single zero with $f(\bk) \approx (k_x - i \sgn(C) k_y)^{|C|}$, and (ii) first-order zero (1FZ) gauge where $f(\bk)$ has $|C|$ first-order zeroes with $f(\bk) \approx k_x - i \sgn(C) k_y$. We note that Ref.~\cite{TurnerOptimalWannier} also discusses how to move vortices through multiplication by suitably chosen meromorphic functions, and has developed an electrostatic picture for their optimal location for Chern band Wannier states.

\emph{Coherent state basis}--- A coherent state basis in a Chern band~\cite{qiGenericWaveFunctionDescription2011,jianCrystalsymmetryPreservingWannier2013, Rashba} is an exponentially localized basis which is overcomplete in the thermodynamic limit due to the Wannier obstruction. If the basis naturally represents translations, there should be $N$ basis elements (for a single band) labelled by lattice sites, which span a vector space of dimension less than $N$ in the limit $N \rightarrow \infty$. 
Coherent states in the LLL are generated by applying lattice magnetic translations to a Gaussian wavefunction ~\cite{Glauber1,Glauber2,PerelomovCompleteness} and are overcomplete by exactly one state due to the Perelomov identity~\cite{PerelomovCompleteness}. 

Coherent states $|\chi_\bR \rangle$ can be constructed via the Fourier transform of $|\chi_\bk \rangle = P_\bk |\tau_\bk \rangle = f(\bk) |\psi_\bk \rangle$, $|\chi_\bR \rangle := \frac{1}{\sqrt{N}} \sum_{\bk \in \BZ} e^{-i \bk \cdot \bR} |\chi_\bk \rangle$. Since $|\chi_\bk \rangle$ is smooth and periodic, its Fourier transform $|\chi_\bR \rangle$ is exponentially localized. If $f(\bk)$ has a zero at $\bk = \bk_0$, then $|\psi_{\bk_0} \rangle$ is orthogonal to  $|\chi_\bR \rangle$ for all $\bR$, leading to 
\begin{equation}
    |\chi_{\bk_0}\rangle = 0 = \frac{1}{\sqrt{N}} \sum_{\bR} e^{i \bk_0 \cdot \bR} |\chi_\bR \rangle
    \label{Overcompleteness}
\end{equation}
which generalizes the Perelemov identity to Chern bands. Thus, if $f(\bk)$ has $n > 0$ zeroes, the coherent states will be overcomplete by at least $n$. Although coherent states are exponentially localized, their localization length cannot be made arbitrarily small due to overcompleteness. This is seen by taking the inner product of (\ref{Overcompleteness}) with $|\chi_{0}\rangle$, leading to the bound
\begin{equation}
    \sum_{\bR \neq 0} e^{i \bk_0 \cdot \bR} \langle \chi_0|\chi_\bR \rangle = -1, \, \implies\, \sum_{\bR \neq 0} |\langle \chi_0|\chi_\bR \rangle| \geq 1
\end{equation}
Note that although the coherent state basis is missing some states, the original Bloch states can be recovered (up to gauge ambiguity) by constructing the projector $P_\bk = \frac{|\chi_\bk \rangle \langle \chi_\bk|}{|f(\bk)|^2}$ away from zeros of $f(\bk)$ and using the smoothness of $P_\bk$ to define it at zero points of $f(\bk)$ via $P_{\bk_0} = \lim_{\bk \rightarrow \bk_0} P_\bk$.

Crucially, the properties of coherent states impose restrictions on {\it any} localized state that can be constructed in the band. To see this, we note that given any localized state $|\tilde \chi \rangle$, we can construct $\tilde f(\bk) = \langle \psi_\bk|\tilde \chi \rangle$ which yield the coherent states $|\tilde \chi_\bR \rangle = \frac{1}{\sqrt{N}} \sum_\bk e^{- i \bk \cdot \bR} \tilde f(\bk) |\psi_\bk \rangle$. This means there is a one-to-one correspondence between coherent states and localized states in a band.

\textit{Coherent state spread---} We now derive a rigorous bound on the spatial spread of coherent states that also provides a bound on any localized state in the band due to the aforementioned correspondence. Consider an arbitrary coherent state $\sum_{\bk} f(\bk) \ket{\psi_\bk}$, where both $f$ and $\ket{\psi}$ are smooth and $f(\bk)$ is normalized such that $\sum_\bk |f(\bk)|^2 = 1$. Write
$\langle w | \br^2 | w \rangle  = \langle w | r_\mu P r_\mu  | w \rangle+\langle w | r_\mu Q r_\mu | w \rangle,$ where $P = \sum_{\bk} |\psi_\bk\rangle\langle \psi_\bk|$ is the band projector and $Q=1-P$. The first term is simplified by noting that the projected position operator acts like a covariant derivative: $P r_\mu \ket{\psi_{\bk}} =  [-i \partial_{\bk_\mu} + A(\bk)] \ket{\psi_\bk}$, leading to $\langle w | r_\mu P r_\mu  | w \rangle = \sum_\bk|(-i \bnabla_\bk - \bA_\bk) f_\bk|^2$. The second term reduces to $\sum_{\bk} |f(\bk)|^2 \Tr g(\bk) $, where $ g_{\mu \nu}(\bk) = \langle \bk |r_\mu Q r_\nu |\bk\rangle$ is the Fubini Study metric. Defining $\Pi = (-i \overline{\partial}_k - \overline{A})$, where $\overline{\partial} = \frac{1}{2}(\partial_{k_x} + i \partial_{k_y})$ and likewise for $\ov{A}$, we get~\cite{claassenPositionmomentumDualityFractional2015,okumaConstructingVortexFunctions2024}
\begin{equation}
\begin{aligned}
    \langle w | \br^2 | w \rangle & = \sum_{\bk} \ov{f(\bk)} \left( \Pi^\dag \Pi + \tr g(\bk) + \Omega(\bk) \right) f(\bk) \\
    & =\sum_{\bk} \ov{f(\bk)} \left( \Pi \Pi^\dag + \tr g(\bk) - \Omega(\bk) \right) f(\bk).
    \end{aligned}
    \label{eq:coherentbounds}
\end{equation}
which is the Hamiltonian of a particle in a periodic potential $\Tr g(\bk)$ and magnetic field $\Omega(\bk)$. Since  $\Tr g - |\Omega| \geq 0$, this expression is lower-bounded by the smallest eigenvalue of $\Pi \Pi^\dag$ ($\Pi^\dag \Pi$), which is nonzero for $C>0$ ($C<0$). This quantity may be interpreted as the minimal cyclotron gap to the first ``momentum space" LL, where $\Omega(\bk)$ is interpreted as a $\bk$-space magnetic field. For ideal bands satisfying $\Tr g = \pm \Omega(\bk)$~\cite{claassenPositionmomentumDualityFractional2015,royBandGeometryFractional2014,parameswaranFractionalQuantumHall2013,ledwithFractionalChernInsulator2020,wangExactLandauLevel2021,liuRecentDevelopmentsFractional2022}, this bound is tight and we can construct the maximally localized coherent state by finding the lowest energy eigenfunction of $\Pi \Pi^\dag$ ($\Pi^\dag \Pi$) for $C>0$ ($C<0$). 

There are a few simpler bounds we may deduce from \eqref{eq:coherentbounds}. First, since $\Pi^\dag \Pi$ 
is positive semidefinite, we have 
$\langle w | \br^2 | w \rangle \geq \min_\bk\left( \tr g(\bk) + |\Omega(\bk)| \right) $. This implies that $ \langle w | \br^2 | w \rangle$ is bounded by both $\min_\bk\tr g(\bk)$ and $2 \min_\bk |\Omega(\bk)|$. Furthermore, for homogeneous $g, \Omega$, we have $\langle w | \br^2 | w \rangle \geq 2C \ell^2$, where $2\pi \ell^{-2}$ is the BZ area. This bound is saturated by the coherent states of the LLL, as well as higher Chern bands constructed from $|C|$ copies of the LLL. 

Note that this bound differs from the corresponding bound for maximally localized Wannier states which contains the momentum space average of the metric; the reason is that Wannier states have $|f(\bk)| = 1$. For a Chern band, Wannier states have infinite $\langle r^2 \rangle$, such that only coherent states can have nontrivial lower bounds.
The minimization over $\bk$ is in fact required for the above bound; there is no general lower bound on $\langle w | \br^2 | w \rangle$ that depends on the Chern number only, or on the average of $\Tr \,g$ (which itself is lower bounded by the Chern number). To see this, we explicitly construct variational coherent states of a class of toy tight-binding models with concentrated Berry curvature whose size is independent of the Chern number.

Consider Bloch wavefunctions where all $\bk$-dependence is concentrated in a parametrically small area of the BZ (the resulting Berry curvature is also concentrated in that region). Choosing $f(\bk)$ to be a smooth function, varying on an order one scale, that is exponentially small over this small region suppresses all of the $\bk$ dependence. The resulting coherent states 
are nearly independent of $C$ since they avoid the region with the concentrated Berry curvature.
Explicitly, we choose 
\begin{equation}
    u_\bk = \frac{1}{\sqrt{1+|s\zeta(\bk)|^{2C}}}\begin{pmatrix} 1 & (s\zeta(\bk))^{C}\end{pmatrix}^T,
    \label{higherCmodel}
\end{equation}
where $\zeta(\bk) = \frac{\partial_k\sigma(k)}{\sigma(k)} - \frac{\ov{k}\ell}{2}$ is periodic but diverges as $1/k$ as $k \rightarrow 0$. Here $k = k_x + i k_y$
and $\sigma(k)$ is the (modified~\cite{haldane2018modular}) Weierstrass sigma function on a hexagonal BZ. The $2\pi C$ phase winding around $k \approx 0$ leads to Chern number $C$. A smooth gauge is obtained through $\tilde u_\bk = \sigma(k)^C u_\bk$, and a parent tight-binding model with exponentially decaying hoppings is given by the Fourier transform of $h(\bk) = 1-\frac{|\tilde u_\bk\rangle\langle \tilde u_\bk |}{\|\tilde u_\bk\|^2}$. For small $s$, the $\bk$ dependence of the wavefunction is concentrated in a small region surrounding the $\Gamma$ point. Then, $\chi_\bk = e^{-|\zeta(\bk)|^2} u_\bk$ is smooth near $\bk \approx 0$ (since $\zeta(\bk)$ diverges at 0) and suppresses the $\bk$ dependence of $u_\bk$ entirely for sufficiently small $s$. The coherent state, corresponding to Fourier transform of $|\chi_\bk\rangle$, is then essentially independent of the Chern number, as shown explicitly in Fig.\ref{fig:Fig1}(b-d). Thus, there can be no lower bound on coherent state widths that is proportional to $C$ or the momentum-averaged Fubini-Study metric.

\emph{Wannier basis}--- A Wannier basis is an orthonormal complete basis that is power-law localized in a Chern band. Below, we show how the Chern number determines its power-law asymptotics. Consider a smooth non-periodic Bloch state $|\psi_\bk \rangle$ and construct $|\tilde \psi_\bk \rangle = \frac{f(\bk)}{|f(\bk)|} |\psi_\bk \rangle$ which is $\bk$-periodic \cite{footnote4} but has a phase singularity at the zeroes of $f(\bk)$.  The Wannier function centered at $\bR = 0$ is $W(\br) = \frac{1}{\sqrt{N}} \sum_{\bk \in \BZ} e^{i \bk \cdot \br} \frac{f(\bk)}{|f(\bk)|} |u_\bk \rangle$. At large distance $\br$, this integral is dominated by the vicinity of zeros of $f(\bk)$. A zero at $\bk_0$ with winding $m$ such that $f(\bk_0 + \bq) \approx (q_x - i \sgn(m) q_y)^{|m|}$ yields the contribution
\begin{equation}
    w_m(\br) = \int \frac{d^2 \bq}{(2\pi)^2} e^{i (\bq \cdot \br - m \varphi_\bq)} = \frac{i^{|m|} |m| e^{-i m \varphi_\br}}{2\pi |\br|^2} 
     \label{WA1}
\end{equation}
In the SZ gauge with a single zero with winding $C$, we get $W^A(\br) \approx \frac{S}{\sqrt{N}} \psi_{\bk_0}(\br) w_C(\br)$, where $S$ is the total area of the system. For the 1FZ gauge, we get a sum of terms each coming from a first order zero, leading to $W^A(\br) \approx \frac{S}{\sqrt{N}} \sqrt{|C|} \phi(\br) w_{\sgn(C)}(\br)$ where $|\phi \rangle = \frac{1}{\sqrt{|C|}} \sum_{i=1}^{|C|} |\psi_{\bk_i} \rangle$. We see that the wavefunction asymptotics follow the Thouless exponent $1/r^2$~\cite{Thouless1984wannier}. Importantly, the prefactor and winding only depends on the Chern number and are insensitive to other band details such as Berry curvature or quantum metric.

{\emph Example 1}--- Consider the Hamiltonian 
\begin{equation}
    H(\bk) = \left( \begin{array}{cc}
        (m - \cos k_x - \cos k_y) & (\sin k_x - i\sin k_y)^{n} \\
        (\sin k_x + i \sin k_y)^{n} & -(m - \cos k_x - \cos k_y)
    \end{array}\right)
    \label{HD1}
\end{equation}
where $n$ is a positive integer.
This Hamiltonian has Chern number $n\, \sgn(m) \Theta(2 - |m|)$, where $\Theta(x)$ is the Heaviside step function. 
Let us start first from the limit $m \rightarrow \infty$, where the Hamiltonian reduces to $\sigma_z$ and the wavefunction for the lower band is $|\psi_\bk \rangle = (0,1)^T$ which we can use as our trial state $|\tau_\bk \rangle$. As we reduce $m$, the wavefunctions get more complicated but retain non-zero overlap with $|\tau_\bk \rangle$. However, as $m$ crosses the topological transition at $m = 2$, we get a $n$-fold band inversion at $\bk = 0$. The resulting $f(\bk)$ has a single zero with winding $n$ at $\bk = 0$ for $0 < m < 2$, leading to the Wannier function and coherent states shown in Fig.~\ref{fig:Fig1}(e,g) for $m = 1$ verifying the asymptotic expression (\ref{WA1}).

\emph{Example 2}--- As an example of several isolated zeros, consider the Hamiltonian 
\begin{equation}
    H(\bk) = \sigma_x \sin k_x + \sigma_y \sin n k_y + (m - \cos k_x - \cos n k_y) \sigma_z
    \label{HD2}
\end{equation}
where $n$ is a positive integer. This Hamiltonian also has the Chern number $n \, \sgn(m) \Theta(2 - |m|)$.
In the $m \gg 2$ limit, we can choose $|\tau_\bk \rangle = (0, 1)^T$ which remains a good trial state till we cross $m=2$ and get $n$ band-inversions at $\bk_l = (0, 2\pi l/n)$, $l = 0,\dots,n-1$. The overlap function $f(\bk)$ has zeros at $\bk_l$ with $f(\bk_l + \bq) \approx \frac{q_x - i n q_y}{2}$ . To bring this into the simpler form $q_x -  i q_y$, we make use of bump functions $B(\bk)$ which are smooth functions with finite support~\cite{SM}. This allows us to modify $|\tau_\bk \rangle$ only in the vicinity of zeros without affecting its smoothness. The choice $\tau_{\bk,2} = 1$, $\tau_{\bk,1} = \sum_{l=1}^{n} B(\bk - \bk_l) \frac{1-n}{1+n} \frac{q_{l,x} + i n q_{l,y}}{2}$, where $\bq_l = \bk - \bk_l$ yields $f(\bk_l + \bq) \approx \frac{n}{1+n} (q_x - i q_y)$ for $m=1$. The resulting coherent states and Wannier functions are shown in Fig.~\ref{fig:Fig1}(f,h) verifying the asymptotic expression (\ref{WA1}).

\emph{Multi-band case}--- Generalizing our formalism to the multiband case is straightforward. For a system of $M$ bands $|\psi_{n,\bk} \rangle$, $n = 1,\dots, M$, construct a trial basis $|\tau_{n,\bk}\rangle$ and define $F_{nm}(\bk) = \langle \psi_{n,\bk}|\tau_{m,\bk} \rangle$. Eq.~\ref{Chernfk} for the Chern number remains valid if we replace $f(\bk)$ by $\det F(\bk)$. Coherent states can be constructed via the Fourier transform of $|\chi_{n,\bk} \rangle = P_\bk |\tau_{n,\bk} \rangle = \sum_m F_{mn}(\bk) |\psi_{m,\bk} \rangle   $. To construct Wannier states, consider the SVD of $F^T(\bk) = V_\bk \hat f(\bk) W_\bk$ where $V_\bk, W_\bk \in {\rm U}(M)$ and $\hat f(\bk)$ is an $M \times M$ diagonal matrix, and take the Fourier transform of the orthonormal basis $U_{nm}(\bk) |\psi_{m,\bk} \rangle$ where $U(\bk) = V_\bk \frac{\hat f(\bk)}{|\hat f(\bk)|} W_\bk$. The Wannier function tail is controlled by the zeros of $\hat f(\bk)$ which can be simplified to have the form $(k_x - i \sgn(\delta_i) k_y)^{|\delta_i|}$. Assuming the $i$-th zero is in the $r$-th component of $\hat f$ yields the expression
\begin{equation}
    W^A_n(\br) = \frac{S}{2\pi \sqrt{N}} \sum_i \frac{i^{|\delta_i|} |\delta_i| e^{-i \delta_i \phi_\br}}{|\br|^2} [V_{\bk_i} \hat \Delta_{r_i} W_{\bk_i}]_{nm} \psi_{m,\bk}(\br)
\end{equation}
where $[\hat \Delta_{r_i}]_{nm} = \delta_{nm} \delta_{n,r_i}$.

\emph{Discussion and conclusion}--- We conclude by discussing some of the implications of our results. First, our results regarding the tails of the Wannier function imply that any real space lattice model built of orthonormal orbitals, e.g. Ref.~\cite{zang2022real}, will involve $1/r^2$-power law interactions whose strength is strongly constrained by the Chern number at long distance. Second, our results place rigorous bounds on {\it any} localized state that can be constructed in a Chern band. This places a lower bound on the minimum size of bound states associated with an impurity potential projected onto the band~\cite{queiroz2024ring} as well as any interacting many-body bound states in the projected band such as Cooper pairs~\cite{Lawanomalous, Lawflat} or excitons~\cite{LawExciton}.

\clearpage
\renewcommand{\theequation}{S\arabic{equation}}
\setcounter{equation}{0}
\renewcommand{\thefigure}{S\arabic{figure}}
\setcounter{figure}{0}
\renewcommand{\thetable}{S\arabic{table}}
\setcounter{table}{0}
\onecolumngrid

\section*{Supplemental Material}
This supplementary material contains detailed discussions on the overlap $f(\bk)$, manipulation of its zeros, and deriving the asymptotics.
\vspace{12 pt}

\section{Manipulating zeros of $f($\texorpdfstring{$\bm{k}$}{k}$)$}
Our analysis is centered on the choice of a basis $|\tau_\bk \rangle$ and using the properties of the overlap function $f(\bk) := \langle \psi_\bk|\tau_\bk \rangle$ to understand the properties of different real space bases of the band. Thus, zero points of $f(\bk)$ play a crucial role in our analysis. Notice that for different $|\tau_\bk\ra$, there are different zero points and the behavior near them is also different. In this section, we consider how to manipulate zero points of $f(\bk)$, both by changing the behavior of $f(\bk)$ near its zero points and by moving those zero points to different positions by choosing appropriate trial states. 

In general, $f(\bk)$ has $m$ zero points $\bk_i, (i=1,\cdots,m)$. The behavior close to the $i$-th zero has the form 
\begin{align} \label{fk}
		f(\bk_i+\bq)\approx (\alpha_{i,1}q+\beta_{i,1}\overline{q})\cdots (\alpha_{i,l_i}q+\beta_{i,l_i}\overline{q})=\prod_{j=1}^{l_i}(\alpha_{i,j}q+\beta_{i,j}\bar q).
\end{align}
Here, $\alpha_{i,j},\beta_{i,j}$ are complex numbers, $q=q_x+i q_y$ is a complexified coordinate, and $l_i$ is the order of zero point $\bk_i$. 
We may prove \eqref{fk} by considering a general expansion $A_{\mu_1, \ldots \mu_m} q_{\mu_1} \ldots q_{\mu_m}$, where $\mu_i = x,y$. Writing this expression in terms of $q$ and $\ov{q}$ yields $\sum_{n=0}^{l_i} A_{n} q^{n} \ov{q}^{l_i-n}$, for some coefficients $A_n$. Pulling out a factor of $|q|^{l_i}$, we obtain $|q|^{l_i} \sum_n A_n e^{2 i n \varphi_\bq}$. This is a polynomial in $z = e^{2 i n \varphi_\bq}$ that can be factored as $c\prod_{i =1}^n (z-z_i)$, which is of the form \eqref{fk} once re-expressed in terms of $q$ and $\ov{q}$.

The winding around a zero $\bk_i$ can be defined through the contribution of that zero to the integral $\frac{1}{2\pi i} \oint d \log f(\bk)$. Assuming $|\alpha_{i,j}| \neq |\beta_{i,j}|$ since otherwise, there exists a direction in $\bq$ where $f$ vanishes, we can perform the integral for a small ball $B_\epsilon$ of radius $\epsilon$ around $\bk = \bk_i$. The integral reduces to a sum of $l_i$ individual terms each of which has the form
\begin{equation}
    \delta_{ij} = \frac{1}{2\pi} \int_0^{2\pi} d\phi \frac{\alpha_{i,j} e^{i \phi} - \beta_{i,j} e^{-i \phi}}{\alpha_{i,j} e^{i \phi} + \beta_{i,j} e^{-i \phi}} = \frac{1}{2\pi i} \oint_{B_1} \frac{dz}{z} \frac{z^2 - (\beta_{i,j}/\alpha_{i,j}) }{z^2 + (\beta_{i,j}/\alpha_{i,j})}
\end{equation}
The $z$-integral is over the unit circle. The integrand has poles at 0 and $\pm i \sqrt{\beta_{i,j}/\alpha_{i,j}}$ with residues $-1$, $+1$ and $+1$ respectively. If $|\beta_{i,j}| > |\alpha_{i,j}|$ then the two latter poles are outside the unit circle and do not contribute thus the integral evaluates to $\delta_{i,j} = -1$. Otherwise, all three poles contribute leading to $\delta_{i,j} = +1$. Thus, $\delta_{i,j} = {\rm sgn}(|\alpha_{i,j}| - |\beta_{i,j}|)$ and the total winding of the $i$-th zero is $\delta_i = \sum_{j=1}^{l_i} \delta_{i,j} = \sum_{j=1}^{l_i} {\rm sgn}(|\alpha_{i,j}| - |\beta_{i,j}|)$. 

We note that the order of a zero is generally different from the absolute value of its winding. For instance, the order and winding number of $q\bar q$ are 2 and 0 respectively; the order and winding number of $q^2$ is 2 and 2. We will now show a method of constructing $|\tau_\bk\ra$ such that behavior near zero point $\bk_i$ is simple, i.e. proportional to $q^{\frac{l_i + \delta_i}{2}} \bar q^{\frac{l_i-\delta_i}{2}}$, and zero points of $f(\bk)$ can be moved to different positions.

\subsection{Changing zero point behavior}
We discuss how to change zero point behavior in Eq.(\ref{fk}) such that it is proportional to $q^{\frac{l_i + \delta_i}{2}} \bar q^{\frac{l_i-\delta_i}{2}}$. Consider an arbitrary smooth and periodic basis $|\tau_\bk \rangle$, we can always apply a smooth and periodic unitary transformation to bring it to the form $|\tau_{\bk}\ra=(1,0,\cdots,0)^T$. Thus, without loss of generality, we can take $|\tau_{\bk}\ra=(1,0,\cdots,0)^T$ leading to $f(\bk)=\la \psi_\bk|\tau_\bk\ra=\psi_{1\bk}^*$, and the asymptotic behavior for $\psi_{1\bk_i+\bq}^*$ is 
\begin{equation} \label{asy1}
	\psi_{1\bk_i+\bq}^*\approx (\alpha_{i,1}q+\beta_{i,1}\overline{q})\cdots (\alpha_{i,l_i}q+\beta_{i,l_i}\overline{q}).
\end{equation}
At $\bk_i$, we know $\psi_{1\bk_i}=0$, then some other components of $|\psi_\bk\ra$ must not be zero because $\la \psi_\bk|\psi_\bk\ra=1$. We assume a non-zero component is $\psi_{a_i\bk}$, i.e. $\psi_{a_i\bk_i}\neq 0$.

Before discussing how to change behavior, we introduce a useful function, a bump function. The motivation is that we only want manipulate behavior near $\bk_i$ and do not want to change $f(\bk)$ globally, and a bump function helps achieve this. A bump function $B(\bk)$ is a real and smooth function on torus $T^2$ that is only non-zero on a open subset $U$ and is 1 on a compact subset $A \subset U$. 

To change behavior near zero points, we can change $\tau_{a_i \bk}$ so that it is not zero any more. For unnormalized $|\tau_\bk\ra$, we choose
\begin{align} \label{tauaik}
	&\tau_{1\bk}=1, \notag\\
	&\tau_{a_i\bk}=\gamma_{i,1}B_i(\bk)g_{i}^{(0)}(\bk)\frac{\overline{\alpha_{i,1}q+\beta_{i,1}\overline{q}}}{\alpha_{i,1}q+\beta_{i,1}\overline{q}}
	=\gamma_{i,1}B_i(\bk)(\overline{\alpha_{i,1}} \overline{q}+\overline{\beta_{i,1}}q)\prod_{j=2}^{l_i} (\alpha_{i,j} q+\beta_{i,j}\overline{q}),
\end{align} 
where $\gamma_{i,1}$ is a complex number to be decided, and $B_{i}(\bk)$ is a bump function that is 1 only around $\bk_i$. Then the behavior of this $f(\bk)$ becomes
\begin{align} \label{ftau1}
	f(\bk_i+\bq)&= \la \psi_{\bk_i+\bq} | \tau_{\bk_i+\bq} \ra \approx (\alpha_{i,j} q+\beta_{i,j}\overline{q}) \left( 1+\psi^*_{a_i\bk_i}\gamma_{i,1}B_i(\bk_i+\bq)\frac{\overline{\alpha_{i,1}} \overline{q}+\overline{\beta_{i,1}}q}{\alpha_{i,1}q+\beta_{i,1}\overline{q}}  \right) \prod_{j=2}^{l_i} (\alpha_{i,j} q+\beta_{i,j}\overline{q}).
\end{align}
By choosing $|\gamma_{i,1}|<1$, we guarantee that $\left|1+\psi^*_{a_i\bk_i}\gamma_{i,1}B_i(\bk_i+\bq)\frac{\overline{\alpha_{i,1}} \overline{q}+\overline{\beta_{i,1}}q}{\alpha_{i,1}q+\beta_{i,1}\overline{q}}\right|>0$, so there is no extra zeros. We can take $B_i(\bk_i+\bq)=1$ when considering behavior near $\bk_i$, so
\begin{align}
	f(\bk_i+\bq)&\approx [(\alpha_{i,1}+\psi_{a_i \bk_i}^*\gamma_{i,1}\ov{\beta_{i,1}})q +(\beta_{i,1}+\psi_{a_i \bk_i}^*\gamma_{i,1}\ov{\alpha_{i,1}})\ov q ]      \prod_{j=2}^{l_i} (\alpha_{i,j} q+\beta_{i,j}\overline{q}).
\end{align}

Now we choose  $\gamma_{i,1}$ to be 
\begin{align} \label{gamma}
	\gamma_{i,1}=\begin{cases}
		-\frac{\beta_{i,1}}{\psi_{a_i\bk_i}^*\ov{\alpha_{i,1}}},\qquad |\alpha_{i,1}|>|\beta_{i,1}|\\
		-\frac{\alpha_{i,1}}{\psi_{a_i\bk_i}^*\ov{\beta_{i,1}}},\qquad |\alpha_{i,1}|<|\beta_{i,1}|,
	\end{cases}
\end{align}
to either vanish as $q$ or $\bar q$. This ensures $|\gamma_{i,1}|<1$. Then $f(\bk_i+\bq)$ becomes 
\begin{equation}
	f(\bk_i+\bq)\approx
	\begin{cases}
		\alpha_{i,1}(1-\frac{|\beta_{i,1}|^2}{|\alpha_{i,1}|^2})q\,\prod_{j=2}^{l_i} (\alpha_{i,j} q+\beta_{i,j}\overline{q}), \qquad |\alpha_{i,1}|>|\beta_{i,1}|,\\
		\beta_{i,1}(1-\frac{|\alpha_{i,1}|^2}{|\beta_{i,1}|^2})\overline{q}\,\prod_{j=2}^{l_i} (\alpha_{i,j} q+\beta_{i,j}\overline{q}), \qquad |\alpha_{i,1}|<|\beta_{i,1}|.
	\end{cases}
\end{equation}

Repeating this procedure for $j=2,\cdots,l_i$, we finally get 
\begin{equation} \label{asy2}
	f(\bk_i+\bq) \approx A_iq^{\frac{l_i+\delta_i}2}\overline{q}^{\frac{l_i-\delta_i}2},
\end{equation}
where $A_i$ is a complex constant. We can further let $A_i$ be real by cancelling the phase of $A_i$ through $|\tau_\bk\ra$.

\subsection{Moving zero points}

In this section we discuss how to construct $|\tau_\bk\ra$ in order to move zero points to different positions. Similar to our earlier discussion, we still assume that $|\tau_{\bk}\ra=(1,0,\cdots,0)^T$ and the behavior of $f(\bk)$ near zero points are Eq.(\ref{asy2}). 
That is to say, $f(\bk)=\psi_{1\bk}^*$ has zero points at $\bk_i,\ (i=1,\cdots,m)$ and zero point behavior is 
\begin{equation} \label{m1}
	f(\bk_i+\bq)=\psi_{1\bk_i+\bq}^* \approx A_iq^{\frac{l_i+\delta_i}2}\overline{q}^{\frac{l_i-\delta_i}2}.
\end{equation}

In order to remove zero point at $\bk_i$, we want $\tau_{1\bk_i+\bq}$ proportional to $\overline{q}^{\frac{l_i+\delta_i}2}q^{\frac{l_i-\delta_i}2}$ which has the opposite winding of $\psi_{1\bk_i+\bq}^*$ so that $\psi_{1\bk_i+\bq}^* \tau_{1\bk_i+\bq}$ is proportional to $|q|^{2l_i}$, and we can add a constant to lift this zero point. Because, $\tau_{1\bk}$ is periodic, it must have other zero points to make sure the total winding number is zero. Therefore, $\tau_{1\bk}$ needs to have other zero points.

We can construct a periodic function $F(\bk)$ such that zero points of $F(\bk)$ are $\bk_1,\cdots,\bk_m, \tilde \bk_1,\cdots,\tilde\bk_n$, the asymptotic behavior for $\bk_i+\bq$, are $F(\bk_i+\bq)\approx c_i\overline{q}^{\frac{l_i+\delta_i}2}q^{\frac{l_i-\delta_i}2}$, and the asymptotic behavior for $\tilde\bk_i+\bq$ is $F(\tilde\bk_i+\bq)\approx \,\tilde c_i  q^{\frac{\tilde l_i +\tilde \delta_i}2} \bar q^{\frac{\tilde l_i -\tilde \delta_i}2}$. Here $c_i$ and $\tilde c_i$ are some non-zero constant parameters. $\tilde \delta_i,\tilde l_i$ are non-negative integers and represent the winding number and order at $\tilde k_i$. We have $\sum_{i=1}^m \delta_i=\sum_{i=1}^{n} \tilde \delta_i$ to ensure the total winding number of $F(\bk)$ vanishes. We will show that $\tilde \bk_i$ are zero points of new $f(\bk)$ and its behavior $F(\tilde\bk_i+\bq)$ is behavior of new $f(\bk)$. That is to say, we can construct $F(\bk)$ to decide zero point positions and behavior of $f(\bk)$.

	Let us choose the unnormalized $|\tau_{\bk}\ra$ to be 
	\begin{align} \label{movetau}
		\tau_{1\bk}=F(\bk)+\sum_{i=1}^{n} \tilde \epsilon_i \tilde B_i(\bk) \psi^*_{\tilde a_i \bk},\qquad\tau_{a\bk}=i\sum_{i=1}^{m}A_ic_i\epsilon_i\psi_{a_i\bk}B_i(\bk)\delta_{a,a_i}-\sum_{i=1}^{n}\tilde \epsilon_i \tilde B_i(\bk) \psi^*_{1 \bk} \delta_{a,\tilde a_i},
	\end{align} where $\epsilon_i,\tilde \epsilon_i\in \mathbb{R}$ are constant, $\tilde B_i(\bk)$ is a bump function near $\tilde\bk_i$, and $\tilde a_i$ is decided similarly as $a_i$, i.e. a component of $\psi_\bk$, $\psi_{\tilde a_i\tilde\bk_i}$, which is non-zero at $\tilde \bk_i$. $F(\bk)$ in $\tau_{1\bk}$ is used to make behavior near $\bk_i$ proportional to $|q|^{2l_i}$, and the first term in $\tau_{a\bk}$, $iA_ic_i^*\epsilon_i\psi_{a_i\bk}B_i(\bk)$, is used to generate a constant number near $\bk_i$ to lift this zero point. The second term in $\tau_{1\bk}$ and $\tau_{a\bk}$, $\tilde \epsilon_i \tilde B_i(\bk) \psi^*_{\tilde a_i \bk}$ and $-\tilde \epsilon_i \tilde B_i(\bk) \psi^*_{1 \bk}$ will cancel each other in $f(\bk)$. They are used to make sure the norm $\tau_\bk$ is non-zero near $\tilde \bk_i$ because $\psi_{\tilde a_i\tilde\bk_i}\neq 0$.

	Now let us check zero points of $f(\bk)$. $f(\bk)$ now becomes
	\begin{equation}
		f(\bk)=F(\bk)\psi_{1\bk}^*+i\sum_{i=1}^{m}A_ic_i^*\epsilon_iB_i(\bk)|\psi_{a_i\bk}|^2.
	\end{equation}
	
		For $ \bk \in \supp (B_i)$ with $i=1,\cdots,m$, we have 
	\begin{align}
		f(\bk_i+\bq)&=F(\bk_i+\bq)\psi_{1\bk_i+\bq}^*+iA_ic_i\epsilon_iB_i(\bk_i+\bq)|\psi_{a_i\bk_i+\bq}|^2\notag\\
		&\approx A_ic_i(|q|^{2l_i}+i\epsilon_i B_{i}(\bk_i+\bq)|\psi_{a_i\bk_i+\bq}|^2).
	\end{align}
	The first term $|q|^{2l_i}$ is real while the second is imaginary, so there is no root. 
	
	For $ \bk \notin \supp (B_i)$, we have $f(\bk)=F(\bk)\psi_{1\bk}^*$ which has zero points at $\tilde \bk_i$ with behavior $f(\tilde\bk_i+\bq)\approx \tilde c_i\psi_{1\tilde\bk_i}^*  q^{\frac{\tilde l_i +\tilde \delta_i}2} \bar q^{\frac{\tilde l_i -\tilde \delta_i}2}$. We see that zero points of $f(\bk)$ and its behavior near zero points are highly related to $F(\bk)$, and we are able to move zero points, cancel zero points with opposite winding and combining zero points by constructing appropriate $F(\bk)$.

	\subsection{Example}
Consider the Dirac model as a concrete example of manipulating zero points of $f(\bk)$ by choosing different trial states $|\tau_\bk\ra$. The model we consider is 
\begin{align}
	H=\sin k_x \sigma_x + \sin (Ck_y) \sigma_y + (1-\cos k_x-\cos (Ck_y))\sigma_z=d_1\sigma_x+d_2\sigma_y+d_3\sigma_z.
\end{align}
Define $d=\sqrt{d_1^2+d_2^2+d_3^2}$ for simplicity. The lower band wavefunction is 
\begin{equation}
	|\psi_{\bk}\rangle = \begin{pmatrix}
		-\sin(\frac \theta 2)\\\ \cos(\frac \theta 2)e^{i\phi}
	\end{pmatrix}
	=
	\begin{pmatrix}
		-\sqrt{\frac{d_1^2+d_2^2}{2d(d+d_3)}}\\ \frac{d_3+d}{\sqrt{2d(d+d_3)}}\frac{d_1+id_2}{\sqrt{d_1^2+d_2^2}}
	\end{pmatrix}.
\end{equation}
We show in this section that we can make zero point behavior of $f(\bk)$ proportional to $q_x \pm i q_y$ and move zero points to different positions by choosing appropriate $|\tau_\bk\ra$.

\subsubsection{simple trial function}
Let's choose the simple trial state $|\tau_{\bk}\ra=(0,1)^T$, then $f(\bk)=\frac{d_3+d}{\sqrt{2d(d+d_3)}}\frac{d_1-id_2}{\sqrt{d_1^2+d_2^2}}$. This has $C$ zero point at $\bk_i=(0,2\pi\frac {i-1}C),\ (i=1,\cdots,C)$ with zero point behavior $f(\bk_i+\bq)\approx \frac{q_x- i C q_y}2$.

We want to make behavior near zero points proportional to $q_x- i q_y$, and after that, move these $C$ zero points to the same place. 

Before starting constructing appropriate $|\tau_\bk\ra$, we introduce two useful functions, a bump function $B(\bk)$ and Weierstrass $\sigma$-function $\sigma(k)$.

(a) Introduce a bump function $B(\bk)$ on torus.,
\begin{equation}
	B(\bk)=\frac{h(2-10|\bk|)}{h(2-10|\bk|)+h(10|\bk|-1)}=\begin{cases}
		0,&|\bk|\geq 0.2\\
		\in(0,1),&0.1<|\bk|<0.2\\
		1,&|\bk|\leq 0.1
	\end{cases},
\end{equation}
where
\begin{equation}
	h(x)=\begin{cases}
		e^{-1/x},&x>0\\
		0,&x\leq 0
	\end{cases}.
\end{equation}
A bump function is smooth on torus, non-zero for $|\bk|<0.2$, and is 1 for $|\bk|<0.1$.

(b) Introduce Weierstrass $\sigma$-function $\sigma(k)$,

	\begin{equation}
    \sigma(k) = \frac{b_1}{\theta'_1(0)} \theta_1\left(\frac{k}{b_1}\Big|\frac{b_2}{b_1} \right) e^{\frac{\bar b_1 \mathcal B}{4 b_1} k^2}, \qquad \theta_1(k|\omega) = \sum_{n = -\infty}^{\infty} e^{i \pi \omega (n+1/2)^2 + 2\pi i (n+1/2)(k + 1/2)},
    \label{sigmaz}
\end{equation}
where $k=k_x+i k_y$ is a comlexified coordinate, $b_1$ and $b_2$ are comlexified reciprocal vectors of Brillouin zone and $\mathcal B= \frac{2\pi}{A_{\rm BZ}}$ (${A_{\rm BZ}}$ is the area of Brillouin zone). Here, $\theta_1(k|\omega)$ is  Jacobi theta function. 

$\sigma(k)$ is a holomorphic function. It has quasi-periodicity that for $G=m b_1+ n b_2,\ (m,n \in \mathbb Z)$, we have $\sigma(k+G)=\sigma(k)e^{\mathcal B \frac{\bar G}2(k+\frac G2)}$. In the first Brillouin zone, $\sigma(k)$ has one zero point $(0,0)$ with behavior near this zero point $\sigma(q)\approx q_x+iq_y$. This means that winding number of $\sigma(k)$ in BZ is 1. 

Multiplying two Weierstrass $\sigma$-functions, we can construct a periodic function that has two zero points with winding number $1$ and $-1$ respectively. The way to do so is defining 
\begin{align}
	v(\bk|\bk_1,\bk_2)=\sigma(k-k_1)\bar\sigma(k-k_2) e^{-\mathcal B\frac{(\bk-\bk_1)^2+(\bk-\bk_2)^2-\bar k (k_2-k_1)+k(\bar k_2-\bar k_1)}4} .
\end{align}
 $v(\bk|\bk_1,\bk_2)$ is a periodic function with two zero points at $\bk_1$ and $\bk_2$. Behavior near $\bk_1$ is $\bar\sigma(\bk_1-\bk_2)e^{-\mathcal B \frac{(\bk_1-\bk_2)^2-\bar k_1 k_2+k_1\bar k_2}{4}}(q_x+iq_y)$, and behavior near $\bk_2$ is $\sigma(k_2-k_1) e^{-\mathcal B \frac{(\bk_1-\bk_2)^2-\bar k_1 k_2+k_1\bar k_2}{4}}(q_x-iq_y)$. $F(\bk)$ for moving zero points of $f(\bk)$ can be constructed from multiplying several $v(\bk|\bk_1,\bk_2)$.
 
 \subsubsection{Change zero point behavior}
 We show how to construct trial states to make behavior near zero points proportional to $q_x -iq_y$.
 Write 
 \begin{align}
 	|\tau_{\bk}^{(1)}\ra =\frac 1{\sqrt{|\tau_{1\bk}^{(1)}|^2+|\tau_{2\bk}^{(1)}|^2}}\begin{pmatrix}
 		\tau_{1\bk}^{(1)}\\ \tau_{2\bk}^{(1)} 
 	\end{pmatrix}.
 \end{align}
Making use of Eq.(\ref{tauaik}) and Eq.(\ref{gamma}), we choose
\begin{align}
	\tau_{1\bk}^{(1)}=\sum_{i=1}^C B(\bk-\bk_i) \frac{1-C}{1+C}\frac{\sin(k_x-k_{ix})+i C\sin((k_y-k_{iy}))}2,\qquad \tau_{2\bk}^{(1)}=1.
\end{align}

Then we find zero point behavior at $\bk_i$ is $(\psi_{1\bk_i}=-1)$ 
\begin{align}
	f(\bk_i+\bq)\approx -\frac{1-C}{1+C} \frac{q_x+iC q_y}2+\frac{q_x-iC q_y}2=\frac C{1+C} (q_x-iq_y).
\end{align}

Mentioned in the paragraph above Eq.(\ref{m1}), for $|\tau^{(1)}_\bk\ra \neq |\tau_\bk\ra=(0,1)^T$, we can perform a unitary transformation $U_\bk^\dagger$ to both $|\psi_\bk\ra$ and $|\tau^{(1)}_\bk\ra$ to make sure $|\tau^{(1)}_\bk\ra=U_\bk|\tau_{\bk}\ra $. Here the unitary matrix is 
\begin{align}
	U=\frac 1{\sqrt{|\tau^{(1)}_{1\bk}|^2+|\tau^{(1)}_{2\bk}|^2}} \begin{pmatrix}
		\tau^{(1)*}_{2\bk} & \tau^{(1)}_{1\bk} \\ -\tau^{(1)*}_{1\bk} & \tau^{(1)}_{2\bk}
	\end{pmatrix}.
\end{align}

Define $|\psi^{(1)}_\bk\ra=U_\bk^\dagger|\psi_{\bk}\ra $ for simplicity,
\begin{align}
	\psi^{(1)}_{1\bk}= \frac {\psi_{1\bk}\tau_{2\bk}^{(1)}-\psi_{2\bk}\tau_{1\bk}^{(1)}}{\sqrt{|\tau^{(1)}_{1\bk}|^2+|\tau^{(1)}_{2\bk}|^2}},\qquad
	\psi^{(1)}_{2\bk}= \frac {\psi_{1\bk}\tau_{1\bk}^{(1)*}+\psi_{2\bk}\tau_{2\bk}^{(1)*}}{\sqrt{|\tau^{(1)}_{1\bk}|^2+|\tau^{(1)}_{2\bk}|^2}}.
\end{align}

We can write $f^{(1)}(\bk)=\la \psi_\bk |\tau^{(1)}_{\bk}\ra=\la \psi_\bk|U|\tau_{\bk}\ra =\la \psi^{(1)}_\bk |\tau_\bk\ra $. This means that we can still let $|\tau_\bk\ra=(0,1)$ but need to consider the transformed wave function $|\psi_\bk^{(1)}\ra$.

\subsubsection{Move zeros}
We discuss how to move zero points. For instance, we want to move zeros to $\tilde \bk$. $\tilde \bk$ can be $(\frac{\pi}2,0)$, $(\frac{\pi}2,\frac{\pi}2)$ or some other point. Write the trial states to be
\begin{align}
	|\tau_\bk^{(2)}\ra =U \frac 1{\sqrt{|\tau_{1\bk}^{(2)}|^2+|\tau_{2\bk}^{(2)}|^2}}\begin{pmatrix}
 		\tau_{1\bk}^{(2)}\\ \tau_{2\bk}^{(2)} 
 	\end{pmatrix}.
\end{align}

Making use of Eq.(\ref{movetau}), we choose

\begin{align}
	\tau^{(2)}_{1\bk}&=\sum_{i=1}^C B(\bk-\bk_i)\ i\bar\sigma(\bk_i-\tilde\bk)e^{-\mathcal B \frac{(\bk_i-\tilde\bk)^2-\bar k_i \tilde k+k_i\bar {\tilde k}}{4}}-B(\bk-\tilde \bk)\psi_{2\bk}^{(1)*},\notag\\
	\tau^{(2)}_{2\bk}&=\prod _{i=1}^Cv(\bk|\bk_i,\tilde \bk) + B(\bk-\tilde \bk)\psi_{1\bk}^{(1)*}.
\end{align}
Here, $F(\bk)=\prod _{i=1}^Cv(\bk|\bk_i,\tilde \bk)$ behaves like $q_x+i q_y$ around $\bk_i$ and $(q_x-i q_y)^C$ around $\tilde\bk$. The first term of $\tau^{(2)}_{1\bk}$, $\sum_{i=1}^C B(\bk-\bk_i)\ i\bar\sigma(\bk_i-\tilde\bk)e^{-\mathcal B \frac{(\bk_i-\tilde\bk)^2-\bar k_i \tilde k+k_i\bar {\tilde k}}{4}}$, is used to lift the zero point $\bk_i$. The second term of $\tau^{(2)}_{1\bk}$ and $\tau^{(2)}_{2\bk}$, i.e. $-B(\bk-\tilde \bk)\psi_{2\bk}^{(1)*}$ and $B(\bk-\tilde \bk)\psi_{1\bk}^{(1)*}$, are used to keep the norm of $|\tau^{(2)}_\bk\ra$ greater than zero. They cancel each other in $f(\bk)$.

Then we see that $f^{(2)}(\bk)=\la\psi_{\bk} | \tau_{\bk}^{(2)}\ra $ has zero point at $\tilde\bk$ and its behavior near $\tilde\bk$ is
\begin{align}
	f(\tilde\bk+\bq)\approx \frac {\psi^{(1)*}_{2\tilde\bk}}{\sqrt{|\tau_{1\tilde\bk}^{(2)}|^2+|\tau_{2\tilde\bk}^{(2)}|^2}} (q_x-iq_y)^C \prod_{i=1}^C \sigma(\tilde\bk -\bk_i)e^{-\mathcal B \frac{(\tilde \bk-\bk_i)^2-\bar \bk_i \tilde\bk +\bk_i\bar{\tilde \bk}}{4}}.
\end{align}

\section{Chern number and zeros of $f($\texorpdfstring{$\bm{k}$}{k}$)$ from periodic gauge}

In this section, we show how to relate the Chern number to the zeros of $f(\bk)$ in a gauge which is periodic but not smooth. Specifically, we show the Chern number is equal to negative total winding number of zeros of $f(\bk)$. Let us assume that $f(\bk)$ has $m$ zero points $\bk_i, (i=1,\cdots,m)$ whose behavior near $\bk_i$ is given by (\ref{fk}) and define the winding of the $i$-th zero as before. Because both $|\psi_\bk\ra$ and $|\tau_\bk\ra$ are periodic, $f(\bk)$ needs to be periodic, so the total winding number of $f(\bk)$ vanishes. This means that there exist several points $\tbk_i$, $(i=1,\cdots,\tilde m)$ such that $f(\tbk_i+\bq)$ has winding number $c_i$ around $\tbk_i$ with $\sum_{i=1}^{\tilde m} \tilde \delta_i= -\sum_{i=1}^{ m} \delta_i$. The periodicity and smoothness of $|\tau_\bk\ra$ implies that $|\psi_\bk\ra$ is singular at points $\tbk_i$ and has winding number $-\tilde\delta_i$ at $\tbk_i$. The minus sign here comes from that $f(\bk)=\la\psi_\bk|\tau_\bk\ra$ which corresponds to the conjugation of $\psi_\bk$. Since the projection operator $P_\bk=\{|\psi_{i\bk}\ra\la \psi_{j\bk}|\}_{i,j}$ is smooth, the components of $|\psi_\bk\ra$ must have the same singular behavior. Therefore, we can extract the commong phase singularity at $\tilde \bk_i$ by writing
\begin{align} \label{psi-sin-def}
	|\psi_{\tbk_i+\bq} \ra \approx e^{i\Theta_i(\bq)} |\xi_{i, \tbk_i+\bq} \ra,\qquad \Theta_i(e^{2\pi i}\bq)- \Theta_i(\bq)=-2\pi \tilde \delta_i,
\end{align}
where $|\xi_{i, \tbk_i+\bq} \ra$ is smooth and normalized. Here, $|u_\bk\ra=e^{-i\bk\cdot \hat\br}|\psi_\bk\ra$, and we see that $|u_\bk\ra$ has the same singular behavior as $|\psi_\bk\ra$ because the factor $e^{-i\bk\cdot \hat\br}$ is smooth. Therefore, we have 
\begin{align} \label{uk-def}
	|u_{\tbk_i+\bq} \ra \approx e^{i\Theta_i(\bq)} |\phi_{i, \tbk_i+\bq} \ra.
\end{align}
where $|\phi_{i, \tbk_i+\bq}\ra=e^{-i(\tbk_i+\bq)\cdot \hat\br}|\xi_{i, \tbk_i+\bq} \ra$ is also smooth and normalized. 

With the information of $|u_\bk\ra$, let us consider the Chern number. The Chern number is defined by
	\begin{align}
		&C=- \frac {1}{2\pi} \int_{T^2} \dm^2 k \mathcal F_{xy},\qquad 
		\mathcal F_{xy} = \partial_{k_x} \mA_y - \partial_{k_y} \mA_x,\qquad
		{\bm{\mA}}= i \la u_\bk | \nabla | u_\bk \ra .
	\end{align}

Since $|u_\bk\ra$ is smooth everywhere except $\tbk_i$, we can separate the Brillouin zone into two regions:  small regions around $\tilde\bk_i$, i.e. $R_1=\cup_i \tBe$ where $\tBe= \{\bk| |\bk-\tbk_i|<\epsilon  \}$, and the other regions on torus, i.e. $R_2=T^2 \setminus R_1$. Here, $\epsilon$ is a small number. Berry connection is smooth in $R_2$ but is singular in $R_1$, so we can write Chern number as
\begin{align}
	C&=-\frac {1}{2\pi} \int_{R_1} \dm^2 k \mathcal F_{xy} - \frac {1}{2\pi} \int_{R_2} \dm^2 k \mathcal F_{xy}, \notag\\
	&= -\frac {1}{2\pi} \int_{R_1} \dm^2 k \mathcal F_{xy} + \frac {1}{2\pi}\sum_{i=1}^{\tilde m}\oint_{\partial \tBe} {\bm{\mA}}_i \cdot \dm \bk,
\end{align}
where ${\bm{\mA}}_i$ is Berry connection around $\tilde \bk_i$. The first term can be ignored in the limit of $\epsilon \rightarrow 0$ due to the finiteness of $\mathcal F_{xy}$. For the second term, we substitute Eq.(\ref{uk-def}) into the above equation and get
\begin{align}
	{\bm{\mA}}_i \approx i \la \phi_{i,\tbk_i+\bq}| \nabla |\phi_{i,\tbk_i+\bq} \ra - \nabla \Theta_i (\bq).
\end{align}

Denote the first term $i \la \phi_{i,\tbk_i+\bq}| \nabla |\phi_{i,\tbk_i+\bq} \ra$ by $\bmA_i^{(\phi)}$. Since $\phi_{i}(\bq)$ is smooth, $\oint_{\partial\tBe} \bmA_i^{(\phi)} \cdot \dm \bk=\int_{\tBe} \dm^2 k\ \mathcal F_{i,xy}^{(\phi)} \rightarrow 0 $ where $\mathcal F_{xy}^{(\phi)} = \partial_{k_y} \mA_x^{(\phi)} - \partial_{k_x} \mA_y^{(\phi)} $ is finite. Therefore, the Chern number is 
\begin{align} \label{chern-delta}
	C&=-\frac {1}{2\pi}\sum_{i=1}^{\tilde m}\oint _{\partial \tBe}\nabla \Theta_i(\bq) \cdot \dm \bq=-\frac {1}{2\pi}\sum_{i=1}^{\tilde m}\Theta_i(e^{2\pi i} \bq )-\Theta_i(\bq) =\sum_{i=1}^{\tilde m} \tilde\delta_i =-\sum_{i=1}^{m} \delta_i.
\end{align}
Therefore, we find that Chern number is equal to negative total zero points winding number of $f(\bk)$.


\begin{thebibliography}{10}
\bibitem{footnote3}
{$\br$ is an arbitrary position in continuum models and a discrete set of
  positions for tight-binding models.}
\bibitem{footnote1}
{For the only if direction, the presumed exponenentially localized Wannier
  states of the band may be used as the trial states $|\tau_\bR \rangle$.}
\bibitem{footnote2}
{Note that since the Hamiltonian itself is periodic, the shifted function can
  only change by a gauge}.
\bibitem{footnote4}
{since $f(\bk)$ transforms in the opposite way under reciprocal lattice
  translations}.
\end{thebibliography}
\end{document}